\documentstyle[12pt, fleqn]{article}
\newcommand{\lr}{LRSM$\,\,$}
\newcommand{\ymh}{Yang-Mills-Higgs lagrangian$\;$}

\newcommand{\nonum}{\nonumber\\}
\newcommand{\be}{\begin{equation}}
\newcommand{\ee}{\end{equation}}
\newcommand{\ba}{\begin{eqnarray}}
\newcommand{\ea}{\end{eqnarray}}
\newcommand{\bd}{\begin{displaymath}}
\newcommand{\ed}{\end{displaymath}}

\renewcommand{\a}{&\!\!\!\!}
\begin{document}
\thispagestyle{empty}
\begin{center}
\vspace*{1.0cm}
{\Large \bf
Reconstruction of the spontaneously broken
\vskip 2mm
gauge theory in non-commutative geometry
}
\vskip 2.5cm
{
\large
Yoshitaka {\sc Okumura}\footnote{
e-mail address: okum@isc.chubu.ac.jp
},
}
\vskip 0.2cm
{\it Department of Natural Science, Chubu University, Kasugai 487, Japan}
\vskip 2mm
and
\vskip 2mm
{
\large
Katsusada {\sc Morita}\footnote{e-mail address: h44753a@nucc.cc.nagoya-u.ac.jp}
}
\vskip 4mm
{\it Department of Physics, Nagoya University, Nagoya 464-01, Japan}
\vskip 2mm
\end{center}
\vspace{2.5 cm}
\begin{abstract}
The scheme previously proposed by the present authors is modified
to incorporate the strong interaction
by affording the direct product internal symmetry.
We do not need to prepare the extra  discrete space
 for the color gauge group responsible for the strong interaction
 to reconstruct the standard model
 and the left-right symmetric gauge model(\lr).
The approach based on non-commutative
geometry leads us to
presents many attractive points such as
the unified picture of the gauge  and Higgs field as the generalized
connection on the discrete space $M_4\times Z_{\mathop{}_{N}}$.
This approach leads us to unified picture of gauge and Higgs fields
as the generalized connection.
 The standard model needs $N=2$ discrete space  for reconstruction
in this formalism.
\lr is still alive as a model with the intermediate symmetry of the
spontaneously broken SO(10) grand unified theory(GUT). $N=3$ discrete space
is needed for the reconstruction of \lr
to include two Higgs bosons $\phi$ and $\xi$
which are as usual transformed as $(2,2^\ast,0)$
and $(1,3,-2)$ under SU(2)$_{\mathop{}_{L}}\times$SU(2)$_{\mathop{}_{R}}
\times$ U(1)$_{\mathop{}_{Y}}$, respectively. $\xi$ is responsible
to make $\nu_{\mathop{}_{R}}$ Majorana fermion
and so well explains the seesaw mechanism.
Up and down quarks have the different masses through the vacuum expectation
value of $\phi$.
\end{abstract}
\vfill\eject
\setlength{\baselineskip}{0.85cm}
\section*{{\large \bf  1. Introduction}}
Since Connes \cite{Con} has proposed
the idea to understand the Higgs mechanism
based on non-commutative geometry (NCG) in the discrete space $M_4\times Z_2$,
several different versions have appeared to make NCG approach more applicable
and understandable, and many works have been done so far
${\cite{Cham}\sim \cite{Naka}}$. \par
One of versions has been initiated by Sitarz \cite{Sita}
in a way  more familiar with the ordinary differential geometry.
The present authors \cite{MO2}
have developed the formalism to be able to apply to
the gauge theory with complex symmetry breaking pattern
by introducing the more fundamental fields $a_i(x,n)$
and the matrix function $M_{nm}$,
where $x$ is the variable in ordinary Minkowski space $M_4$,
$i$ is a variable in the extra internal space
and $n,m$ are the arguments in the discrete space $Z_{\mathop{}_{N}}$.
Gauge fields are composed of $a_i(x,n)$, which is similar to the effective
gauge field in Berry phase \cite{Ber}. This may provide a key
to explore the extra internal space with the variable $i$.
The matrix $M_{nm}$ determines the scale and pattern of the spontaneous
breakdown of gauge symmetry and composes the Higgs field
along with $a_i(x,n)$.
Thus, in our formalism Higgs fields are created at the same time
as the spontaneous breakdown of symmetry takes place, which leads
to the quite different viewpoint of Higgs mechanism from the ordinary one.
Then, the unified picture of the gauge fields and Higgs fields
as the generalized
connection in NCG on the discrete space $M_4\times Z_{\mathop{}_{N}}$ is
realized.\par
According to our formalism \cite{MO2} we could reconstruct the
standard model \cite{MO1}, \cite{MO2}, SU(5) grand unified theory (GUT)
\cite{MO2}
and SO(10) GUT \cite{O10}.
The color SU(3) gauge group responsible for the strong interaction is
well incorporated in the reconstructions of SU(5) and SO(10) GUTs
from the outset.
However, in the case of the standard model \cite{MO1}, \cite{MO2}
the extra sheet of the discrete
space is needed to incorporate the color SU(3) gauge group,
which makes the reconstruction of the standard model rather awkward.
\par
The purpose of this paper is to modify the previous formalism \cite{MO2}
without introducing the extra discrete space for the strong interaction and
to incorporate the color SU(3) gauge
field(gluons)
by affording the direct product internal symmetry
\hfill\break (semi-simple group) such as
SU(3)$_c\times$SU(2)$_{\mathop{}_{L}}$.
The standard model needs $N=2$ discrete space  for reconstruction
in this formalism, while it needed $N=3$ discrete space
in the previous formalism \cite{MO1}, \cite{MO2}.
\lr is still alive as a model with the intermediate symmetry of the
spontaneously broken SO(10) grand unified theory(GUT). $N=3$ discrete space
is needed for the reconstruction of \lr
to include two Higgs bosons $\phi$ and $\xi$
which are as usual transformed as $(2,2^\ast,0)$
and $(1,3,-2)$ under
SU(2)$_{\mathop{}_{L}}\times$SU(2)$_{\mathop{}_{R}}\times$U(1)$_{\mathop{}_{Y}}
$, respectively.
$\xi$ is responsible
for the first stage breakdown of symmetry to give rise to the seesaw mechanism
and so make $\nu_{\mathop{}_{R}}$ Majorana fermion .
Up and down quarks have the different masses from each other
through the vacuum expectation value of $\phi$ leaving U(1)$_{em}$ invariant.
\par
This paper is divided into five sections. The next section presents
the modifications of our previous formalism
based on the generalized differential calculus on
$M_4\times Z_{\mathop{}_{N}}$. The generalized gauge field is defined there
and a geometrical
picture for the unification of the gauge and Higgs fields is realized.
The third section is the application to the standard model
which leads to rather clear scheme compared with our previous one.
The fourth section  provides the reconstruction of \lr.
It will be realized that
the boson and leptonic sectors are nicely incorporated
so that the seesaw mechanism
works well to give the right-handed neutrino huge mass and the left-handed
neutrino extremely small mass.
 The last section is devoted for concluding remarks.
\noindent
\section*{
\large \bf  2. Generalized gauge field with direct product form}
The generalized gauge field $ A(x,n)$ in non-commutative geometry
on the discrete space $M_4\times Z_{\mathop{}_{N}}$
was given in Ref.\cite{MO2} as
\ba
&&      A(x,n)=\sum_{i}a^\dagger_{i}(x,n){\bf d}a_i(x,n), \label{2.1}
\ea
where $a_i(x,n)$ is the square-matrix-valued function
and ${\bf d}$ is the generalized exterior derivative defined as follows.
\ba
      && {\bf d}a_{i}(x,n)=(d + d_\chi)a(x,n)
                                   =(d+\sum_{k=1}^{\mathop{}_{N}}
                                     d_{\chi_k})a(x,n) \nonum
  &&   da_i(x,n) = \partial_\mu a_i(x,n)dx^\mu,\hskip 1cm \nonum
  && d_{\chi_k} a_i(x,n) = [-a_i(x,n)M_{nk} + M_{nk}a_i(x,k)]\chi_k.
  \label{2.2}
\ea
Here
$dx^\mu$ is
ordinary one form basis, taken to be dimensionless, in $M_4$, and $\chi_k$
is the one form basis, assumed to be also dimensionless,
in the discrete space $Z_{\mathop{}_{N}}$ with the variable
$n\,(n=1,\cdots,N)$.
We have introduced $x$-independent matrix $M_{nm}$ whose hermitian
conjugation is given by $M_{nk}^\dagger=M_{kn}$.
The matrix $M_{nk}$ turns out to determine the scale
and pattern of the spontaneous breakdown of the gauge symmetry.
We here skip to explain detailed algebras with respect to
non-commutative geometry because those are seen in Ref.\cite{MO2}.
According to Ref.\cite{MO2}, we can define
the gauge fields $A_\mu(x,n)$ and the Higgs field $\Phi_{nm}(x)$ as
\ba
&&    A_\mu(x,n) = \sum_{i}a_{i}^\dagger(x,n)\partial_\mu a_{i}(x,n), \nonum
&&\Phi_{nm}(x) = \sum_{i}a_{i}^\dagger(x,n)\,(-a_i(x,n)M_{nm}
            + M_{nm}a_i(x,m)).\label{2.3}
\ea
\par
We extend Eq.(\ref{2.1})
to the generalized gauge field with the direct product
form to incorporate gluon field on the same sheet as flavor gauge fields.
\be
      {\cal A}(x,n)=\sum_{i}a^\dagger_{i}(x,n){\bf d}a_i(x,n)\otimes 1
      +1 \otimes \sum_{j}b^\dagger_{j}(x,n){\bf d}b_j(x,n),\label{2.4}
\ee
where the second term is responsible for the gluon field,
 so that actually
 ${\bf d}b_j(x,n)={d}b_j(x)$ because the strong interaction does not
 break down spontaneously. We assume $b_j(x,n)=b_j(x)$
 which means the strong interaction
 works on the every discrete space $(n=1,2,\cdots,N)$.
In the same context as in Eq.(\ref{2.3}), the gluon field $G^{}_\mu(x)$
is expressed as
\be
   G^{}_\mu(x)=\sum_{j}b_{j}^\dagger(x)\partial_\mu b_{j}(x),\label{2.5}
\ee
where $n$-independence of $G^{}_\mu(x)$ means that the gluon field
exists on the every sheet ($n=1,2,\cdots N)$ as stated above.
In order to identify  $A_\mu(x,n)$ and $G^{}_\mu(x)$ as true gauge fields,
the following conditions have to be imposed.
\ba
&&    \sum_{i}a_{i}^\dagger(x,n)a_{i}(x,n)= 1,  \nonum
&&     \sum_{j}b_{j}^\dagger(x)b_{j}(x)={1\over g_s},
 \label{2.6}
\ea
where $g_s$ is a constant related to
the corresponding coupling constant shown later. $i$ is a variable
of the extra
internal space which we can not now identify what it is. Eqs.(\ref{2.3})
and (\ref{2.5})
are very similar to the effective gauge field
in Berry phase \cite{Ber}, which may
lead to the identification of this internal space.
In general, we can put the right hand side of the first equation
in Eq.(\ref{2.6})
to be ${1\over g_n}$. However, we put it as it is to avoid the complexity.
\par
Before constructing the gauge covariant field strength,
we address the gauge transformation of $a_i(x,n)$ and $b_j(x)$
which is defined as
\ba
&&      a^{g}_{i}(x,n)= a_{i}(x,n)g(x,n), \nonum
&&      b^{g}_{j}(x) =  b_j(x)g_s(x),
\label{2.7}
\ea
where
$g(x,n)$ and $g_s(x)$ are the gauge functions with respect to
the corresponding
flavor unitary group and the color SU(3)$_c$ group, respectively.
Then, we can get the gauge transformation of ${\cal A}(x,n)$ to be
\ba
{\cal A}^g(x,n)\a =\a g^{-1}(x,n)\otimes g^{-1}_s(x){\cal A}(x,n)g(x,n)\otimes
g_s(x)\nonum
&&+g^{-1}(x,n){\bf d}g(x,n)\otimes 1+1\otimes {1\over g_s}
\,g^{-1}_s(x){d}g_s(x),
  \label{2.8}
\ea
where use has been made of Eq.(\ref{2.6}) and  as in Eq.(\ref{2.2}),
\ba
       {\bf d}g(x,n)\a=\a
      \partial_\mu g(x,n)dx^\mu \nonum
                &&+ [-g(x,n)M_{nk} + M_{nk}g(x,k)]\chi_k.
        \label{2.9}
\ea
Eq.(\ref{2.8}) affords us
to construct the gauge covariant field strength as follows:
\be
  {\cal F}(x,n)= F(x,n)\times 1 + 1 \times {\cal G}(x),
\label{2.10}
\ee
where $F(x,n)$ and ${\cal G}(x)$ are the field strengths of flavor
and color gauge fields, respectively and given as
\ba
&&     F(x,n) = {\bf d}A(x,n)+A(x,n)\wedge A(x,n)     \nonum
&&     {\cal G}(x)   =d\,G^{}(x)+g_s G^{}(x)\wedge G^{}(x).
\label{2.11}
\ea
The algebras of non-commutative differential geometry defined
in Ref.\cite{MO2} yields
\ba
 F(x,n) \a=\a { 1 \over 2}F_{\mu\nu}(x,n)dx^\mu \wedge dx^\nu +
               \sum_{m\ne n}D_\mu\Phi_{nm}(x)dx^\mu \wedge \chi_m \nonum
              && + \sum_{k\ne n}V_{nk}(x)\chi_k \wedge \chi_n \nonum
     && + \sum_{k\ne n}\sum_{\{m\ne k,n\}}V_{nkm}(x)\chi_k \wedge \chi_m,
                \label{2.12}
\ea
where
\ba
&&  F_{\mu\nu}(x,n)=\partial_\mu A_\nu (x,n) - \partial_\nu A_\mu (x,n)
               + [A_\mu(x,n), A_\mu(x,n)],\nonum
&&  D_\mu\Phi_{nm}(x)=\partial_\mu \Phi_{nm}(x) -
  (\Phi_{nm}(x)+M_{nm})A_\mu(x,m)\nonum
&&         \hskip 3.5cm + A_\mu(x,n)(M_{nm} + \Phi_{nm}(x))\nonum
&&  V_{nk}(x)= (\Phi_{nk}(x) + M_{nk})(\Phi_{kn}(x) + M_{kn}) -
             Y_{nk}(x),\hskip 0.5cm {\rm for}\,\,k\ne n\nonum
&&  V_{nkm}(x)= (\Phi_{nk}(x) + M_{nk})(\Phi_{km}(x) + M_{km}) -
             Y_{nkm}(x). \hskip 0.2cm{\rm for}\,\,k\ne n,  m\ne n
     \label{2.13}
\ea
$Y_{nk}(x,n)$ and $Y_{nlm}$ in Eq.(\ref{2.13})
are auxiliary fields and expressed as
\ba
&&  Y_{nk}(x)= \sum_{i}a_{i}^\dagger(x,n)M_{nk}M_{kn}a_{i}(x,n),
  \hskip 1cm {\rm for} \,\,n\ne k,\nonum
&&  Y_{nkm}(x)= \sum_{i}a_{i}^\dagger(x,n)M_{nk}M_{km}a_{i}(x,m).
  \hskip 1cm {\rm for} \,\,n\ne k, m,\;k\ne m
 \label{2.14}
\ea
If we define $H_{nm}(x) = \Phi_{nm}(x) + M_{nm},$
it is readily known that
the function $H_{nm}(x)$ represents the unshifted Higgs field,
whereas $\Phi_{nm}(x)$ denotes the shifted Higgs field
with vanishing vacuum expectation value so
that $M_{nm}$ determines the scale and pattern of the
spontaneous breakdown of gauge symmetry.
In contrast to $F(x,n)$, ${\cal G}(x)$ is simply denoted as
\ba
    {\cal G}(x)\a =\a {1\over 2}{\cal G}_{\mu\nu}(x)dx^\mu\wedge dx^\nu \nonum
        \a =\a {1\over 2}\{\partial_\mu G^{}_\nu(x)-\partial_\nu G^{}_\mu(x)
        + g_s[G^{}_\mu(x), G^{}_\mu(x)]\}dx^\mu\wedge dx^\nu
\label{2.15}
\ea
\par
With the same metric structure on the discrete space
$M_4\times Z_{\mathop{}_{N}}$ as in Ref.\cite{MO2}
 we can obtain the gauge invariant
\ymh(YMH)
\ba
{\cal L}_{{\mathop{}_{YMH}}}(x)\a\a\!\!\!=
-\sum_{n=1}^{\mathop{}_{N}}{1 \over g_{n}^2}
< {\cal F}(x,n),  {\cal F}(x,n)>\nonum
\a\a\!\!\!= -{\rm tr}\sum_{n=1}^{\mathop{}_{N}}{1\over 2g^2_n}
F_{\mu\nu}^{\dag}(x,n)F^{\mu\nu}(x,n)\nonum
\a\a\!\!\!+{\rm tr}\sum_{n=1}^{\mathop{}_{N}}
\sum_{m\ne n}{\alpha^2\over g_{n}^2}
    (D_\mu\Phi_{nm}(x))^{\dag}D^\mu\Phi_{nm}(x)\nonum
-{\rm tr}\sum_{n=1}^{\mathop{}_{N}}
{\alpha^4\over g_{n}^2}\a\a\!\!\!\!\sum_{k\ne n} V_{nk}^{\dag}(x)
V_{nk}(x)
-{\rm tr}\sum_{n=1}^{\mathop{}_{N}}{\alpha^4\over g_{n}^2}
         \sum_{k\ne n}\sum_{m\ne n,k}V_{nkm}^{\dag}(x)V_{nkm}(x),\nonum
&&-{\rm tr}\sum_{n=1}^N{1\over 2g_{n}^2}{\cal G}_{\mu\nu}^{\dag}(x)
{\cal G}^{\mu\nu}(x),
\label{2.16}
\ea
where $g_n$ is a constant relating
to the coupling constant of the flavor gauge field and
tr denotes the trace over internal symmetry matrices. $\alpha$ emerges from
the definition of metric $<\chi_n, \chi_m>=-\alpha^2 \delta_{nm}$.
The third term is the potential term of Higgs particle and the fourth term
is the interaction term between Higgs particles.
\par
Let us turn to the fermion sector to construct the Dirac Lagrangian. This is
also deeply indebted to Ref.\cite{MO2}
so that only main points should be explained
by skipping details. Let us start to define the covariant derivative acting
on the spinor field $\psi(x,n)$ which is the representation
of the corresponding semi simple group including SU(3)$_c$.
\be
{\cal D}\psi(x,n)=({\bf d}+ {\cal A}^f(x,n))\psi(x,n),
\label{2.17}
\ee
which we call the covariant spinor one-form.
The algebraic rules in Ref.\cite{MO2} along with Eq.(\ref{2.4})
leads Eq.(\ref{2.17}) to
\ba
{\cal D}\psi(x,n)\a=\a \{1\otimes 1\partial_\mu
+ (A^f_\mu(x,n)\otimes 1dx^\mu+\sum_m H_{nm}(x)\otimes 1\chi_m) \nonum
 \a\a+1\otimes G^{f}_\mu(x)dx^\mu\}\psi(x,n),
\label{2.18}
\ea
where $A^f_\mu(x,n)$ and $G^{f}_\mu(x)$ are the differential representations
with respect to $\psi(x,n)$ and do not  necessarily coincide with $A_\mu(x,n)$
and $G^{}_\mu(x)$ in the gauge boson sector, respectively.
It should be noticed
that ${\cal D}\psi(x,n)$ is gauge covariant so that
\be
     {\cal D}\psi^g(x,n)=(g^f(x,n))^{-1}\otimes
     {(g^f_s(x))}^{-1}{\cal D}\psi(x,n), \label{2.19}
\ee
where
$g^f(x,n)\otimes g^f_s(x)$ is the gauge transformation function
with respect to the representation of $\psi(x,n)$.
Corresponding Eq.(\ref{2.17}), the associated spinor one-form is introduced by
\be
{\tilde {\cal D}}\psi(x,n)= 1\otimes 1\{\gamma_\mu \psi(x,n)dx^\mu
              -i{c_{ \mathop{}_{Y}}}\psi(x,n)\sum_m\chi_m\},
\label{2.20}
\ee
where $c_{ \mathop{}_{Y}}$ is a real dimensionless constant related to the
Yukawa coupling constant between Higgs field and fermions. With the same inner
products for spinor one-forms as in Ref.\cite{MO2}, we can get the Dirac
Lagrangian.
\ba
{\cal L}_{ \mathop{}_{D}}(x,n)\a=\a i<{\tilde {\cal D}}\psi(x,n),
{\cal D}\psi(x,n)>\nonum
\a=\a i\,{\rm tr}\,[\,{\bar\psi}(x,n)\gamma^\mu(1\otimes1\partial_\mu
+A_\mu^{f}(x,n)\otimes 1+ 1\otimes G^{f}_\mu(x))\psi(x,n) \nonum
\a\a+i{g_{ \mathop{}_{Y}}}{\bar\psi}(x,n)\sum_m H_{nm}(x)\otimes1\psi(x,m)\,],
\label{2.21}
\ea
where $c_{\mathop{}_{Y}}$ is changed into ${g_{ \mathop{}_{Y}}}$
by appropriate
replacement. The total Dirac Lagrangian is the sum over $n$:
\be
{\cal L}_{\mathop{}_{D}}(x)
      =\sum_{n=1}^{\mathop{}_{N}}{\cal L}_{\mathop{}_{D}}(x,n),
\label{2.22}
\ee
which is apparently invariant for the Lorentz and gauge transformations.
Eqs.(\ref{2.16}) and (\ref{2.22}) along with Eq.(\ref{2.21})
are crucially important
to reconstruct the
spontaneously broken gauge theory.
\par
With these preparations, we can apply the direct product formalism proposed
in this section to the standard model and the left-right symmetric model, which
make the presentations rather simpler than those before.
\noindent
\section*{
\large \bf  3. Standard model}
Reconstruction of the standard model in non-commutative geometry
on the discrete space has been already completed in Ref.\cite{MO1}, \cite{MO2}
where the extra sheet for the color gauge (gluon) field
is arranged $(N=3)$, which made the presentation rather awkward.
However, the present formulation which affords the incorporation
of direct product gauge fields makes it clear,
because it needs only two sheets $(N=2)$.
\par
We identify $G^{}_\mu(x)$ in Eq.(\ref{2.5}) with the gluon field
as
\be
    G^{}_\mu(x) = -{i\over 2}\sum_{a=1}^8 \lambda^a G_\mu^a(x),
\label{3.23}
\ee
where $\lambda^a$ $(a=1,2,\cdots 8)$ are the Gell-Mann matrices
and
$A_\mu(x,n)$, $\Phi_{nk}(x)$ and $M_{nk}$ $(n,m=1,2)$ as
\ba
&&A_\mu(x,1)=-{i \over 2}\sum_{i=1}^3\tau^i A^i_\mu(x)-{i \over 2}a
\tau^0B_\mu(x),\nonum
&&\Phi_{12}(x)=
\left(
\matrix{
\phi_+(x)\cr
\phi_0(x)\cr
}\right), \hskip 1cm
M_{12}=
\left(
\matrix{
0\cr
\mu\cr
}\right),
\label{3.24}
\ea
where $A_\mu^i(x)$ is $SU(2)$ gauge field and $B_\mu(x)$ is $U(1)$ gauge field,
and
\ba
&&A_\mu(x,2)=-{i \over 2}bB_\mu(x),\nonum
&&\Phi_{21}(x)=\Phi^{\dag}_{12}(x), \hskip 1cm
M_{21}=(0,\;\; \mu)=M^{\dag}_{12}.
\label{3.25}
\ea
$\mu$ in $M_{nk}$ is a real, positive constant
and there exist the free real parameters $a,b$
in Eqs.(\ref{3.24}) and (\ref{3.25}).
\par
{}From these specifications, the generalized field strength ${\cal F}(x,n)$ is
expressed as
\ba
 {\cal F}(x,1)\a=\a -{i \over 4}\sum_a 1\otimes\lambda^a {\cal G}
 _{\mu\nu}^a(x) dx^\mu \wedge dx^\nu,\nonum
\a\a+ {i\over 4}[\,- \sum_i\tau^i F_{\mu\nu}^i(x)
          -a\tau^0 B_{\mu\nu}(x)\, ]\otimes 1dx^\mu \wedge dx^\nu\nonum
               \a+\a  D_\mu H(x)\otimes 1 dx^\mu \wedge \chi_2
         +[\,H(x)H^\dagger(x)-Y_{23}(x)\,]\otimes 1\chi_1\wedge\chi_2,
\label{3.26}
\ea
\ba
  F(x,2)\a=\a -{i \over 4}\sum_a 1\otimes\lambda^a {\cal G}_{\mu\nu}^a(x)
  dx^\mu \wedge dx^\nu-b{i\over 4}B_{\mu\nu}(x)1\otimes 1 dx^\mu \wedge
  dx^\nu \nonum
            \a\a+(\,D_\mu H(x)\,)^\dagger\otimes 1 dx^\mu \wedge \chi_1
      + [\,H^\dagger(x)H(x)-Y_{32}(x)\,]\otimes 1 \chi_2\wedge\chi_1,
    \label{3.27}
\ea
where
\ba
&& {\cal }{\cal G}_{\mu\nu}^a(x)=
      \partial_\mu G_\nu^a(x)-\partial_\nu G_\mu^a(x)
         +g_sf^{abc}G_\mu^b(x)G_\nu^c(x),  \nonum
&&      F_{\mu\nu}^i(x)=\partial_\mu A_\nu^i(x)-\partial_\nu A_\mu^i(x)
         +\epsilon^{ijk}A_\mu^j(x)A_\nu^k(x),  \nonum
&&      B_{\mu\nu}(x)=\partial_\mu B_\nu(x)-\partial_\nu B_\mu(x),\nonum
&&      H(x)=\Phi_{23}(x)+M_{23}.\nonum
&&      D^\mu H(x)=[\,\partial_\mu-{i\over 2}\,(\sum_i\tau^iA_\mu^i(x)
          +(a-b)\,\tau^0\,B_\mu(x)\,)\,]\,H(x).
       \label{3.28}
\ea
{}From here we can proceed in the same way as in Ref.\cite{MO2}
including the handling
of the auxiliary fields $Y_{nk}(x)$ and scale transformations of fields,
so that we skip all of procedures to obtain YMH in the standard model
(See Ref.\cite{MO2}).
\par
Let us turn to the fermionic sector.
 We can achieve the same results as in Ref.\cite{MO2}
 by replacing $\psi(x,2)$ and
 $\psi(x,3)$ in Ref.\cite{MO2} by $\psi(x,1)$ and $\psi(x,2)$, respectively.
As for the leptons, they are singlet for the color gauge group,
so that the differential representation $G^{f}_\mu(x)$ in Eq.(\ref{2.18})
vanishes.
We have the two kind of  specifications $\psi(x,n)(n=1,2)$
 for quark sector to give masses not only down quark but also up quark
 for which the differential representations do not vanish.
  We can carry all calculations in the same way as in Ref.\cite{MO2}
   and get the same
  conclusions.
Thus, we skip all detailed explanations about the fermionic sector.
\noindent
\section*{
\large \bf  4. Left-right symmetric model}
The purpose of this section is to reconstruct the left-right symmetric gauge
model (\lr) in our formalism.
\lr is the gauge theory with the symmetry
SU(3)$_c\times$SU(2)${\mathop{}_{L}}\times\!$SU(2)$_{\mathop{}_{R}}\!
\times$  U$(1)$
and still alive as a model with the intermediate symmetry of the
spontaneously broken SO(10) GUT which is the most promising grand
unified theory in particle physics.
Thus, it is still worthwhile to reconstruct \lr, and
it is also important for the test of our formalism.
Let us begin with the bosonic sector.\par
\centerline{\bf A. Yang-Mills-Higgs sector}
\par
In order to get YMH in \lr we prepare  $(N=3)$ discrete space
with three sheets.\par
Let us begin to specify ${\cal A}(x,n)(n=1,2,3)$.
\ba
     {\cal A}(x,1)=A_{\mu}(x,1)\otimes1+1\otimes G^{}_\mu(x)dx^\mu
      + \Phi_{12}(x)\otimes1\chi_2,
       \label{4.29}
\ea
where
\ba
    &&G^{}_\mu(x)= -{i\over 2}\sum_{a=1}^8 \lambda^a {G}_\mu^a(x),\nonum
    &&A_{\mu}(x,1)=
  -{i \over 2}\sum_k\tau_k A_{{\mathop{}_{L}}\mu}^k-{i \over 2}a\tau^0B_\mu
             =-{i \over 2}
\left(
\matrix{
 A_{{\mathop{}_{L}}\mu}^3+aB_\mu &
             \sqrt{2}W_{{\mathop{}_{L}}\mu} \cr
    \sqrt{2}W_{{\mathop{}_{L}}\mu}^\dagger&
       -A_{{\mathop{}_{L}}\mu}^3+aB_\mu \cr
}\right),
        \label{4.30}
\ea
with the Pauli matrices $\tau_k(k=1,2,3)$, $2\times2$ unit matrix $\tau^0$
  and
$W_{\mathop{}_{L}}=(A_{{\mathop{}_{L}}}^1 -
iA_{{\mathop{}_{L}}}^2)/\sqrt{2}$,
and
\be
\Phi_{12}=\Phi_{21}^\dagger=\phi=\left(
\matrix{
         \phi_2^{0} & \phi_1^+ \cr
         \phi_2^-   & \phi_1^0 \cr
}
\right),
\hskip 0.5cm
M_{12}=M_{21}^\dagger =\left(
\matrix{
  \mu_2 & 0 \cr
  0   & \mu_1 \cr
}\right).
\label{4.31}
\ee
\be
     {\cal A}(x,2))=(A_{\mu}(x,2)\otimes1+1\otimes G^{}_\mu(x))dx^\mu
      + \Phi_{21}(x)\otimes1\chi_1+\Phi_{23}(x)\otimes 1\chi_3,
       \label{4.32}
\ee
where
\be
    A_{\mu}(x,2)=
  -{i \over 2}\sum_k\tau_k A_{{\mathop{}_{R}}\mu}^k-{i \over 2}b\tau^0B_\mu
             =-{i \over 2}
\left(
\matrix{
 A_{{\mathop{}_{R}}\mu}^3+bB_\mu &
             \sqrt{2}W_{{\mathop{}_{R}}\mu} \cr
    \sqrt{2}W_{{\mathop{}_{R}}\mu}^\dagger&
       -A_{{\mathop{}_{R}}\mu}^3+bB_\mu \cr
}\right),
        \label{4.33}
\ee
with
$W_{\mathop{}_{R}}=(A_{{\mathop{}_{R}}}^1 -
iA_{{\mathop{}_{R}}}^2)/\sqrt{2}$,
and
\be
\Phi_{23}=\Phi_{32}^\dagger=\xi=
\left(
\matrix{
 \xi^{-} &    \xi^{0} \cr
  \xi^{--} & - \xi^{-} \cr
}\right),
\hskip 0.5cm
M_{23}=M_{32}^\dagger =
\left(
\matrix{
      0 & M \cr
      0  & 0 \cr
}\right).
\label{4.34}
\ee
\be
     {\cal A}(x,3)=(A_{\mu}(x,3)\otimes1+1\otimes G^{}_\mu(x))dx^\mu
      +\Phi_{32}(x)\otimes\chi_2,
       \label{4.35}
\ee
where
\be
    A_{\mu}(x,3)=
  -{i \over 2}\sum_k\tau_k A_{{\mathop{}_{R}}\mu}^k-{i \over 2}c\tau^0B_\mu
             =-{i \over 2}
\left(
\matrix{
 A_{{\mathop{}_{R}}\mu}^3+cB_\mu &
             \sqrt{2}W_{{\mathop{}_{R}}\mu} \cr
    \sqrt{2}W_{{\mathop{}_{R}}\mu}^\dagger&
       -A_{{\mathop{}_{R}}\mu}^3+cB_\mu \cr
}\right).
        \label{4.36}
\ee
We assume $M_{13}=M_{31}^\dagger=0$ and so $\Phi_{13}=\Phi_{31}=0$
in the above specifications. There are three parameters $a,b$ and $c$
in above equations. However, $a=b$ is required because the hypercharge
of $\phi$ is zero.
We have to take gauge transformation functions as
\ba
&&g\,(x,1)=e^{-ia\alpha(x)}g_{\mathop{}_{L}}(x),\;e^{-ia\alpha(x)}\in U(1),\;
g_{\mathop{}_{L}}(x)\in SU(2)_{\mathop{}_{L}}, \nonum
&&g\,(x,2)=e^{-ib\alpha(x)}g_{\mathop{}_{R}}(x),\;e^{-ib\alpha(x)}\in U(1),\;
g_{\mathop{}_{R}}(x)\in SU(2)_{\mathop{}_{R}}, \nonum
&&g\,(x,3)=e^{-ic\alpha(x)}g_{\mathop{}_{R}}(x),\;e^{-ic\alpha(x)}\in U(1),\;
g_{\mathop{}_{R}}(x)\in SU(2)_{\mathop{}_{R}}.
 \label{4.37}
\ea
$\phi$ and $\xi$ transform under SU(2)${\mathop{}_{L}} \times
$SU(2)$_{\mathop{}_{R}} \times$U$(1)$  according to $(2,2^*,0)$ and
$(1,3,-2)$ representations, respectively because of $a=b=-c$.
\par
With these assignments we can calculate \ymh ${\cal L}_{{\mathop{}_{YMH}}}(x)$
in Eq.(\ref{2.16}). ${\cal L}_{{\mathop{}_{YMH}}}(x)$ is denoted as
\be
 {\cal L}_{{\mathop{}_{YMH}}}(x) = {\cal L}_{\mathop{}_{GB}}
 + {\cal L}_{\mathop{}_{HK}}
    - { V}_{\mathop{}_{HP}}, \label{4.38}
\ee
where
${\cal L}_{\mathop{}_{GB}}$, ${\cal L}_{\mathop{}_{HK}}$ and
${ V}_{\mathop{}_{HP}}$  are
the gauge boson kinetic term,  the Higgs boson kinetic term, and
the Higgs boson potential term respectively.
\ba
{\cal L}_{\mathop{}_{GB}}\a=\a-{1\over 4}
  \left(\sum_{i=1}^3{1\over g_i^2}\right) \sum_a G_{\mu\nu}^a G^{a\mu\nu}\nonum
\a\a-{1\over 4}
{1\over g_1^2}\sum_i F^i_{{\mathop{}_{L}}\mu\nu} F_{\mathop{}_{L}}^{i\mu\nu}
-{1\over 4}\left({1\over g_2^2}+{1\over g_3^3}\right)
\sum_i F^i_{{\mathop{}_{R}}\mu\nu} F_{\mathop{}_{R}}^{i\mu\nu}\nonum
\a\a-{1\over4}
\left({a^2\over {g_1^2}}+{b^2\over {g_2^2}}
+{c^2\over {g_3^2}}  \right)B_{\mu\nu}B^{\mu\nu}.
\label{4.39}
\ea
By denoting  $g_0=\left(\sum_{i=1}^3{1\over g_i^2}\right)^{-{1\over2}}$,
$g_{\mathop{}_{L}}=g_1,$
$ g_{\mathop{}_{R}}=\left({{1\over g_2^2}+{1\over g_3^2}}
\right)^{-{1\over2}}$ and
\hfill\break
$ g_{\mathop{}_{B}}=\left(
{a^2\over {g_{\mathop{}_{1}}^2}}+{b^2\over {g_{\mathop{}_{2}}^2}}
+{c^2\over {g_3^2}}\right)^{-{1\over2}}$,
and replacing $G_\mu^a$,
$A_{\mathop{}_{L}\mu}$, $A_{\mathop{}_{R}\mu}$ and $B_\mu$ by
$g_0G_\mu^a$,
$ g_{\mathop{}_{L}}A_{\mathop{}_{L}\mu}$,
$g_{\mathop{}_{R}} A_{\mathop{}_{R}\mu}$ and $g_{\mathop{}_{B}}B_\mu$
, respectively we can find the standard gauge boson kinetic term as
\ba
{\cal L}_{\mathop{}_{GB}}\a=\a-{1\over 4}\sum_a G_{\mu\nu}^a G^{a\mu\nu}\nonum
\a\a -{1\over 4}
\sum_i F^i_{{\mathop{}_{L}}\mu\nu} F_{\mathop{}_{L}}^{i\mu\nu}
-{1\over 4}
\sum_i F^i_{{\mathop{}_{R}}\mu\nu} F_{\mathop{}_{R}}^{i\mu\nu}
-{1\over4}B_{\mu\nu}B^{\mu\nu},
\label{4.40}
\ea
where
\ba
&& G_{\mu\nu}^a  = \partial_\mu G_\nu^a-\partial_\nu G_\mu^a
       +g_sf^{abc}G_\mu^b G_\nu^c, \nonum
&& F_{{(\mathop{}_{L,R})}\mu\nu}^i  = \partial_\mu A_{{(\mathop{}_{L,R})}\nu}^i
          -\partial_\nu A_{{(\mathop{}_{L,R})}\mu}^i
       +g_{\mathop{}_{L,R}}\,\epsilon^{ijk}A_{{(\mathop{}_{L,R})}\mu}^j
       A_{{(\mathop{}_{L,R})
       }\nu}^k,        \nonum
&& B_{\mu\nu}=\partial_\mu B_\nu-\partial_\nu B_\mu,
\label{4.41}
\ea
with the replacement $g_0g_s\rightarrow g_s$.\par
Higgs kinetic term ${\cal L}_{\mathop{}_{HK}}$ is written as follows:
\ba
 {\cal L}_{\mathop{}_{HK}}
       \a=\a {\rm Tr}\left| \partial_\mu\phi-{i\over 2}(g_{\mathop{}_{L}}
   \sum_{k=1}^3 \tau_k A_{\mathop{}_{L\mu}}^k+ag_{\mathop{}_{B}}B_\mu\tau_0)
   (\phi+M_{12})\right.  \nonum
\a\a  \hskip 4cm +\left.{i\over 2}(\phi+M_{12})
   (g_{\mathop{}_{R}}
   \sum_{k=1}^3 \tau_k A_{\mathop{}_{R\mu}}^k+bg_{\mathop{}_{B}}B_\mu\tau_0)
   \right|^2\nonum
     \a+\a {\rm Tr}\left| \partial_\mu\xi-{i\over 2}(g_{\mathop{}_{R}}
   \sum_{k=1}^3 \tau_k A_{\mathop{}_{R\mu}}^k+bg_{\mathop{}_{B}}B_\mu\tau_0)
   (\xi+M_{23})\right.  \nonum
\a\a  \hskip 4cm +\left.{i\over 2}(\xi+M_{23})
   (g_{\mathop{}_{R}}
   \sum_{k=1}^3 \tau_k A_{\mathop{}_{R\mu}}^k+cg_{\mathop{}_{B}}B_\mu\tau_0)
   \right|^2
\label{4.42}
\ea
where the replacements
\ba
&&   \left({\alpha^2 \over g_{\mathop{}_{L}}^2}
     +{\alpha^2 \over g_2^2}\right)^{1\over 2}\phi \rightarrow \phi,
          \hskip 1cm
             \left({\alpha^2\over g_{\mathop{}_{L}}^2}
  +{\alpha^2\over g_{\mathop{}_{2}}^2}\right)^{1\over 2}M_{12} \rightarrow
         M_{12},
     \nonum
&&   \left({\alpha^2\over g_{\mathop{}_{2}}^2}
      +{\alpha^2\over g_{3}^2}\right)^{1\over 2}\xi\rightarrow \xi,
         \hskip 1cm
    \left({\alpha^2\over g_{\mathop{}_{2}}^2}
    +{\alpha^2\over g_{3}^2}\right)^{1\over 2}M_{23}\rightarrow M_{23},
     \label{4.43}
\ea
have been done.\par
Let us move to the Higgs potential terms. In order to avoid the complexity
in expressions, $g_1=g_2=g_3=g$ is assumed without loss of essence.
\ba
{ V}_{\mathop{}_{HP}}\a=
        \a  {g^2 \over 4}{\rm Tr}\left|(\phi+M_{12})(\phi^\dagger+M_{21})-
          Y_{121} \right|^2
         + {g^2 \over 4}{\rm Tr}\left|(\phi^\dagger+M_{21})(\phi+M_{12})-
        Y_{212}  \right|^2 \nonum
         \a+\a{g^2 \over 4}{\rm Tr}\left|(\xi+M_{23})(\xi^\dagger+M_{32})-
           Y_{232} \right|^2
    +{g^2 \over 4}{\rm Tr}\left|(\xi^\dagger+M_{32})(\xi+M_{23})-
           Y_{323} \right|^2 \nonum
      \a + \a  {g^2\over 2}{\rm Tr}\left|(\phi+M_{12})(\xi+M_{23})-
          Y_{123} \right|^2
            \label{4.44}
\ea
where the auxiliary fields $Y_{121}$, $Y_{323}$ and $Y_{123}$
 are independent fields so that
the terms containing these fields in Eq.(\ref{4.44}) vanish due to the equation
of
motion, however,
\ba
 && Y_{212}=
      \sum_ia_i^\dagger(2)\left(\matrix{\mu_2^2 & 0 \cr
                 0              & \mu_1^2 \cr }\right) a_i(2), \nonum
 && Y_{232}=
          \sum_ia_i^\dagger(2)\left(\matrix{ M^2 & 0 \cr
                                0 & 0 \cr}\right)a_i(2),
 \label{4.45}
\ea
have the relation that $M^2(Y_{212}-\mu_1^2)=(\mu_2^2-\mu_1^2)Y_{232}$ so that
Eq.(\ref{4.44}) leads to
\ba
{V}_{\mathop{}_{HP}}\a=\a
 {g^2 \over 2}
 {1\over M^4+(\mu_1^2-\mu_2^2)^2}
 {\rm Tr}\left|M^2\left\{(\phi^\dagger+M_{21})
                  (\phi+M_{12})-M_{21}M_{12}\right\}\right.\nonum
        & &\hskip 2cm  +\left.(\mu_1^2-\mu_2^2)
         \left\{(\xi+M_{23})(\xi^\dagger+M_{32})-M_{23}M_{32}\right\}
                \right|^2.
            \label{4.46}
\ea
The present formulation in non-commutative geometry is very restrictive to
 the Higgs potential and Higgs interacting terms. Eq.(\ref{4.46}) includes
those terms in more restrictive way. We think Eq.(\ref{4.46}) to be taken
at a specially chosen renormalization point. If $\mu_1>\mu_2$,
we can chose the true vacuum
at a point that the vacuum expectation values of $\phi$ and $\xi$ are zero,
where the Higgs potential gets the minimum value so that the spontaneous
symmetry breakdown takes place.
\par
Gauge boson mass matrix can be extracted from Eq.(\ref{4.42}). Thinking of the
neutral
gauge boson in the flavor sector, the remaining gauge symmetry is only
U(1)$_{em}$, which should leads to
the vanishing determinant of the neutral gauge boson mass matrix.
Also for this purpose we have to take the relation $a=b$
which is due to the vanishing hypercharge of $\phi$.
As a result
the following mass matrix for the neutral gauge boson  follows.
\be
M_{\mathop{}_{NGB}}=\left(\matrix{
       g_{\mathop{}_{L}}^2(\mu_1^2+\mu_2^2) &
        -g_{\mathop{}_{L}}g_{\mathop{}_{R}}(\mu_1^2+\mu_2^2)  &   0   \cr
       -g_{\mathop{}_{L}}g_{\mathop{}_{R}}(\mu_1^2+\mu_2^2)   &
       g_{\mathop{}_{R}}^2(\mu_1^2+\mu_2^2+4M^2) &
      2g_{\mathop{}_{R}}g_{\mathop{}_{B}}(a-c)M^2   \cr
      0   &   2g_{\mathop{}_{R}}g_{\mathop{}_{B}}(a-c)M^2
      & g_{\mathop{}_{B}}^2(a-c)^2M^2\cr}\right),
      \label{4.47}
\ee
which surely yields the vanishing determinant.
On the other hand, charged gauge boson mass matrix
is given also from Eq.(\ref{4.42}) as
\be
M_{\mathop{}_{CGB}}=\frac12\left(\matrix{
       g_{\mathop{}_{L}}^2(\mu_1^2+\mu_2^2) &
         -2g_{\mathop{}_{L}}g_{\mathop{}_{R}}\mu_1\mu_2          \cr
       -2g_{\mathop{}_{L}}g_{\mathop{}_{R}}\mu_1\mu_2       &
       g_{\mathop{}_{R}}^2(\mu_1^2+\mu_2^2+M^2) \cr
      }\right). \label{4.48}
\ee
Equations (\ref{4.47}) and (\ref{4.48}) enable us to estimate the gauge boson
masses
in the limit of $M >\!\!> \mu_1,\;\mu_2 $.
According to Eq.(\ref{4.47}), the neutral gauge boson masses are written as
\ba
&&         M_\gamma^2=0, \hskip 1cm \nonum
&&         M_{\mathop{}_{Z}}^2=
(g_{\mathop{}_{R}}^2+g_{\mathop{}_{L}}^2)\left( 1
         +\frac12{((a-c)^2g_{\mathop{}_{B}}^2-4g_{\mathop{}_{R}}^2)\over
     4g_{\mathop{}_{R}}^2+(a-c)^2g_{\mathop{}_{B}}^2}\right)
     (\mu_1^2+\mu_2^2),\nonum
&&         M_{Z_{\mathop{}_{R}}}^2=
         (4{g_{\mathop{}_{R}}^2+(a-c)^2{g_{\mathop{}_{B}}^2}})M^2,
\label{4.49}
\ea
where
$\gamma$ and  $Z$ represent of course photon and neutral
 weak boson, respectively
and ${Z_{\mathop{}_{R}}}$ denotes the extra neutral gauge boson
expected in this model. $M_{Z_{\mathop{}_{R}}}$ is estimated to be so
large that one can not detect ${Z_{\mathop{}_{R}}}$ in the energy range of
the accelerator available nowadays.\par
{}From Eq.(\ref{4.41})
we can find mass matrix of the charged gauge boson masses as
\be
     M_{\mathop{}_{W}}^2=\frac12g_{\mathop{}_{L}}^2(\mu_1^2+\mu_2^2),
     \hskip 2cm
     M_{W_{\mathop{}_{R}}}^2=\frac12g_{\mathop{}_{R}}^2(\mu_1^2+\mu_2^2+2M^2),
\label{4.50}
\ee
where $W$ denotes the charged weak boson and
$W_{\mathop{}_{R}}$ is also the extra charged gauge boson expected in this
model and its mass is so high that it can not be detectable.\par
If we assume $g_{\mathop{}_{L}}^2=g_{\mathop{}_{R}}^2=g$, the Weinberg angle
is determined by
\be
    \sin^2\theta_{\mathop{}_{W}}=1-{M_{\mathop{}_{W}}^2 \over
                  M_{\mathop{}_{Z}}^2 }={\mu_2^2\over \mu_1^2+\mu_2^2}.
\label{4.51}
\ee

\par
\centerline {B. \bf Leptonic sector}
\par
It should be noted that the leptonic sector must be well
designed so that the seesaw mechanism  nicely works to give
the right-handed neutrino the huge mass and the left-handed neutrino
the extremely small mass \cite{Gell}.
$\Psi(x,n)$ is identified for each sheet $n=1,2,3$ as follows:
\ba
&&          \psi(x,1)={1\over\sqrt{2}}l_{\mathop{}_{L}}
           ={1\over\sqrt{2}}\left(\matrix{ \nu_{\mathop{}_{L}}
            \cr
                              e_{\mathop{}_{L}}\cr}\right),
                              \hskip 1cm Y=-1,    \nonum
&&          \psi(x,2)={1\over\sqrt{3}}l_{\mathop{}_{R}}
                      ={1\over\sqrt{3}} \left(\matrix{ \nu_{\mathop{}_{R}} \cr
                              e_{\mathop{}_{R}}   \cr}\right),
                              \hskip 1cm Y=-1,\nonum
&&          \psi(x,3)={1\over\sqrt{3}}l_{\mathop{}_{R}}^c
                      ={1\over\sqrt{3}} \left(\matrix{ e_{\mathop{}_{R}}^c \cr
                                -\nu_{\mathop{}_{R}}^c \cr}\right),
                              \hskip 1cm Y=1,
 \label{4.52}
\ea
where the superscript $c$ denotes the charge conjugation of
$e_{\mathop{}_{R}}$ and $\nu_{\mathop{}_{R}}$.
The Nishijima-Gell-Mann law
$Q=T_{\mathop{}_{L}}^3+T_{\mathop{}_{R}}^3+Y/2$ well works in Eq.(\ref{4.52}).
\hfill\break
\quad From Eqs.(\ref{2.19}),(\ref{2.20}) and (\ref{4.52}),
we determine the covariant spinor one form
${\cal D}\psi(x,n)$ and the associated spinor one form
${\tilde {\cal D}}\psi(x,n)$ to obtain the Dirac Lagrangian for lepton sector,
 where it should be noticed that the differential representation $G^{f}(x)$
 vanishes and $a(=b)$ and $c$ are identified with the hypercharges of the
 corresponding fermions.
\ba
 {\cal D}\psi(x,1)\a=
 \a \{\partial_\mu-{i\over 2}(g_{\mathop{}_{L}}
     \sum_k\tau_kA_{\mathop{}_{L\mu}}^k-g_{\mathop{}_{B}}\tau^0B_\mu)\}
      \psi(1)dx^\mu     \nonum
   \a\a+g_\phi(\phi+M_{12})\psi(2)\chi_2 \nonum
{\tilde {\cal D}}\psi(x,1)\a=\a\gamma_\mu\psi(1)dx^\mu
                           +ic_{_l}\psi(1)\sum_{m\ne 1}\chi_m,
 \label{4.53}
\ea
\ba
 {\cal D}\psi(x,2)\a=\a \{\partial_\mu-{i\over 2}(g_{\mathop{}_{R}}
     \sum_k\tau_kA_{\mathop{}_{R\mu}}^k-g_{\mathop{}_{B}}\tau^0B_\mu)\}
      \psi(2)dx^\mu     \nonum
   \a\a+g_\phi(\phi^\dagger+M_{21})\psi(1)\chi_1
       +g_{\xi}(\xi+M_{23})\psi(3)\chi_3, \nonum
{\tilde {\cal D}}\psi(x,2)\a=\a\gamma_\mu\psi(2)dx^\mu
                           +ic_{_l}\psi(2)\sum_{m\ne 2}\chi_m,
 \label{4.54}
\ea
\ba
 {\cal D}\psi(x,3)\a=\a\{\partial_\mu-{i\over 2}(g_{\mathop{}_{R}}
     \sum_k\tau_kA_{\mathop{}_{R\mu}}^k+g_{\mathop{}_{B}}\tau^0B_\mu)\}
      \psi(3)dx^\mu     \nonum
      \a\a+g_{\xi}(\xi^\dagger+M_{32})\psi(2)\chi_2, \nonum
{\tilde {\cal D}}\psi(x,3)\a=\a\gamma_\mu\psi(3)dx^\mu
                           +ic_{_l}\psi(3)\sum_{m\ne 3}\chi_m.
 \label{4.55}
\ea
{}From these equations together with Eqs.(\ref{2.21}) and (\ref{2.22})
we can get the
Dirac Lagrangian for lepton sector.
\ba
{\cal L}_{\mathop{}_{D}}\a=
\a i{\bar l}_{\mathop{}_{L}}\gamma^\mu
\{\partial_\mu-{i\over 2}(g_{\mathop{}_{L}}
     \sum_k\tau_kA_{\mathop{}_{L\mu}}^k-g_{\mathop{}_{B}}\tau^0B_\mu)\}
     l_{\mathop{}_{L}}\nonum
\a\a +i{\bar l}_{\mathop{}_{R}}\gamma^\mu
\{\partial_\mu-{i\over 2}(g_{\mathop{}_{R}}
     \sum_k\tau_kA_{\mathop{}_{R\mu}}^k-g_{\mathop{}_{B}}\tau^0B_\mu)\}
     l_{\mathop{}_{R}}\nonum
\a\a -g_{\phi}\{{\bar l}_{\mathop{}_{L}}(\phi+M_{12})l_{\mathop{}_{R}}+
       {\bar l}_{\mathop{}_{R}}(\phi^\dagger+M_{21})l_{\mathop{}_{L}}\}\nonum
\a\a -g_{\xi}\{{\bar l}_{\mathop{}_{R}}(\xi+M_{23})l_{\mathop{}_{R}}^c+
    {\bar l}_{\mathop{}_{R}}^c(\xi^\dagger+M_{32})l_{\mathop{}_{R}}\},
\label{4.56}
\ea
where
the Yukawa coupling constants
are changed in scale  appropriately.
Let us put attention on the mass terms in Eq.(\ref{4.56}).
\be
{\cal L}_{\rm lepton\ mass}=
-\mu_1({\bar e}_{\mathop{}_{R}}{e}_{\mathop{}_{L}}
                  +{\bar e}_{\mathop{}_{L}}{e}_{\mathop{}_{R}})-
\mu_2({\bar \nu}_{\mathop{}_{R}}{\nu}_{\mathop{}_{L}}
                  +{\bar \nu}_{\mathop{}_{L}}{\nu}_{\mathop{}_{R}})
                  +M({\bar \nu}_{\mathop{}_{R}}^c{\nu}_{\mathop{}_{R}}
                  +{\bar \nu}_{\mathop{}_{R}}{\nu}_{\mathop{}_{R}}^c),
  \label{4.57}
\ee
where we neglect the coupling constants without loss of generality
and leads to the neutrino mass matrix
\be
      m_\nu=
      \left(\matrix{ 0         & \mu_2 \cr
                     \mu_2 &  -M    \cr}\right),
\label{4.58}
\ee
which is sufficient to give rise to the seesaw mechanism.
$-$ sign of one eigenvalue  of matrix $m_\nu$
is absorbed into the phase of the wave function.
\par
\vskip 0.3cm
\centerline{\bf C. Quark sector}
\par
In contrast to lepton,
quark interacts with gluon via the strong interaction and the  mixing
among three generations is conspicuous in quark sector.
It is not so difficult to  taken into  account the quark mixing, which
prescriptions have been already exhibited in Ref.\cite{MO2}.
Thus, for the sake of brevity, we consider here only one generation.
$\psi(x,n)$ is identified for each sheet $n=1,2,3$ as follows:
\ba
&&          \psi(x,1)={1\over\sqrt{2}}q_{\mathop{}_{L}}
           ={1\over\sqrt{2}}\left(\matrix{ u_{\mathop{}_{L}}
            \cr
                              d_{\mathop{}_{L}}   \cr}\right),
         \hskip 1cm Y={1\over 3},    \nonum
&&          \psi(x,2)={1\over\sqrt{2}}q_{\mathop{}_{R}}
                      ={1\over\sqrt{2}} \left(\matrix{ u_{\mathop{}_{R}} \cr
                              d_{\mathop{}_{R}}   \cr}\right),
          \hskip 1cm Y={1\over 3},\nonum
&&          \psi(x,3)=0,
 \label{4.59}
\ea
where we abbreviate subscripts about triplet representation of $SU(3)_c$.
$\psi(x,3)$ have to be chosen to be zero owing to the respect of
U(1) gauge transformation.
$\psi(x,1)$ are transformed as $(3, 2, 0, {1\over3})$ under
SU(3)$_c\times$SU(2)$_{\mathop{}_{L}}\times$SU(2)$_{\mathop{}_{R}}\times$U(1)
and $\psi(x,2)$  as $(3, 0, 2, {1\over3})$.
As a result $ {\cal D}\psi(x,n)$ and ${\tilde {\cal D}}\psi(x,n)$ are
determined to be
\ba
 {\cal D}\psi(x,1)\a=\a \{1\otimes1\partial_\mu-{i\over 2}g_s\sum_{a=1}^8
 1\otimes\lambda_aG_\mu^a
 -{i\over 2}(g_{\mathop{}_{L}}\sum_k\tau_k\otimes1 A_{\mathop{}_{L\mu}}^k
 \nonum
     \a\a+{1\over 3}g_{\mathop{}_{B}}\tau^0\otimes1 B_\mu)\}\psi(1)dx^\mu
   +g_\phi(\phi +M_{12})\otimes1 \psi(2)\chi_2 \nonum
{\tilde {\cal D}}\psi(x,1)\a=\a\gamma_\mu\psi(1)dx^\mu
                           +ic_q\psi(1)\sum_{m\ne 1}\chi_m,
 \label{4.60}
\ea
\ba
 {\cal D}\psi(x,2)\a=\a\{1\otimes1 \partial_\mu-{i\over 2}g_s\sum_{a=1}^8
  1\otimes\lambda_aG_\mu^a
 -{i\over 2}(g_{\mathop{}_{R}}
     \sum_k\tau_k\otimes1A_{\mathop{}_{R\mu}}^k \nonum
    \a\a +{1\over 3}g_{\mathop{}_{B}}
     \tau^0\otimes1 B_\mu)\}      \psi(2)dx^\mu
   +g_\phi(\phi^\dagger +M_{21})\otimes1 \psi(1)\chi_1, \nonum
{\tilde {\cal D}}\psi(x,2)\a=\a \gamma_\mu\psi(2)dx^\mu
                           +ic_q\psi(2)\sum_{m\ne 2}\chi_m,
 \label{4.61}
\ea
{}From these equations we can obtain the Dirac Lagrangian for quark sector
in first generation as follows:
\be
     {\cal L}^q={\cal L}_{kin}^q+ {\cal L}_{\rm Yukawa}^q,  \label{4.62}
\ee
where
\ba
        {\cal L}_{kin}^q \a=\a
      i{\bar q}_{\mathop{}_{L}}\gamma^\mu\left(
      \partial_\mu-{i\over 2}g_s\sum_{a=1}^8\lambda_aG_\mu^a
              -{i\over 2}g_{\mathop{}_{L}}
\sum_{k=1}^3\tau_kA_{\mathop{}_{L\mu}}^k-{i\over6}g_{\mathop{}_{B}}\tau^0B_\mu
\right)q_{\mathop{}_{L}} \nonum
     \a+\a i{\bar q}_{\mathop{}_{R}}\gamma^\mu\left(
      \partial_\mu-{i\over 2}g_s\sum_{a=1}^8\lambda_aG_\mu^a
              -{i\over 2}g_{\mathop{}_{R}}
\sum_{k=1}^3\tau_kA_{\mathop{}_{R\mu}}^k-{i\over6}g_{\mathop{}_{B}}\tau^0B_\mu
\right)q_{\mathop{}_{R}}
 \label{4.63}
\ea
 and
\be
{\cal L}_{\rm Yukawa}^q=
-g_{\phi}^q\{{\bar q}_{\mathop{}_{L}}(\phi+M_{12})q_{\mathop{}_{R}}+
       {\bar q}_{\mathop{}_{R}}(\phi^\dagger+M_{21})q_{\mathop{}_{L}}\}
\label{4.64}
\ee
with the appropriately denoted coupling constants.
Eq.(\ref{4.64}) yields quark masses that $m_u=g_\phi^q\mu_2$ for up-quark
and $m_d=g_\phi^q\mu_1$ for down-quark.\par
\vskip 0.5cm
\noindent
\section*{
\large \bf  5. Concluding remarks}
\par
We have modified the scheme previously proposed by the present authors to
be able to incorporate the color gauge field(gluon) without the extra discrete
space for it. The modified scheme can afford the direct product gauge group
(semi-simple group) such as SU(3)$_c\times$SU(2)$_{\mathop{}_{L,R}}$ so that
the strong interaction works in every sheet ($n=1,2,\cdots .N)$
of the discrete space $M_4\times Z_{\mathop{}_{N}}$. It should be noticed that
Lagrangian becomes meaningful after being summed over $n$
because such a procedure makes it Hermitian, and so we can not detect the
discrete space $Z_{\mathop{}_{N}}$ experimentally.\par
Then, we have constructed the standard model
and the left-right symmetric gauge theory
in non-commutative geometry using the formalism developed here.
The reconstruction of the standard model need only $N=2$ discrete space, which
makes it rather clearer compared to the previous one \cite{MO1}, \cite{MO2}.
\lr~is still alive as a model with the intermediate symmetry of the
spontaneously broken SO(10) GUT which is the most promising unified theory
in particle physics. Thus, whether we can actually reconstruct \lr
is an important step
to achieve the justification of our formalism. We could nicely explain
the seesaw mechanism to make the right-handed neutrino $\nu_{\mathop{}_{N}}$
Majorana fermion and up and down quarks are given the different masses
from each other. These are the necessary conditions for \lr to be physically
meaningful model. The only controversial point in our reconstruction of \lr
is about the Higgs potential.
In general, the present formulation imposes
the severe constraints on Higgs potential and interacting terms
responsible for the symmetry breakdown.
These constraints worked very well in the reconstruction of SU(5) GUT
owing to the symmetry breakdown due to adjoint and ${\bf 5}$ Higgs
fields in SU(5) representation to result in the nice forms
of Higgs potential and interacting terms \cite{MO2}.
Eq.(\ref{4.46}) includes Higgs potential in more restrictive way.
In this point we should understand that our scheme in NCG gives
 \ymh at a specially chosen renormalization point so that Eq.(\ref{4.46}) takes
 the special form.
 Even so we can chose the true vacuum in Eq.(\ref{4.46})
at a point that the vacuum expectation values of $\phi$ and $\xi$ are zero,
where the Higgs potential gets the minimum value. Therefore,
the spontaneous breakdown of symmetry can take place in the proper way.
 \par
We may discuss the evolution of the coupling constants in the Higgs potential
in Eq.(\ref{4.46}) by use of the renormalization equation.
This subject including the quantization of the gauge theory with
the restricted Higgs potential due to the non-commutative geometry
will be studied in the future work.
\vskip 0.5cm
\centerline{\bf Acknowledgements}
\vskip 0.5cm
\par
One of the authors(Y.O.) would like to
express his sincere thanks to
Professor Y.~Takahashi and Professor H.~Umezawa
for hospitality at University of Alberta. This work was completed
while Y.O. stayed at University of Alberta.
\vskip 0.5cm
\vfill\eject
\def\jmp{J.~Math.~Phys.$\,$}
\def\pl{Phys. Lett.$\,$ }
\def\np{Nucl. Phys.$\,$}
\def\ptp{Prog. Theor. Phys.$\,$}
\def\prl{Phys. Rev. Lett.$\,$}
\def\pr{Phys. Rev. D$\,$}
\def\mp{Int. Journ. Mod. Phys.$\,$ }

\end{document}